\documentstyle[12pt,a4]{article}
\newcommand{\Vev}[1]{\langle 0\vert #1 \vert 0\rangle}
\newcommand{\dnn}{{\bar{\vphantom\i d}}_n}              
\newcommand{\BC}[2]{{ #1 \choose #2}}   
\newcommand{\RE}{\mbox{Re$\:$}}
\newcommand{\IM}{\mbox{Im$\:$}}
\newcommand{\bi}{\bibitem}

\newcommand{\gr}{\overline{g}(t)}
\newcommand{\al}{\overline{\alpha}_s (q^2)}
\newcommand{\vtil}{\tilde{v}}

\begin{document}
\thispagestyle{empty}
\halign to\textwidth{\hfil # \hfil\tabskip 0in
plus \textwidth & \hfil # \tabskip0pt\cr
{\sl Institute of Physics, Czechosl.\ Acad.\ of Sci.}&
{\bf PRA-HEP-92/17}\cr
{\sl and}& {\bf September 1992}\cr
{\sl Nuclear Centre, Charles University}&\cr
{\sl Prague}&\cr}
\vskip 1in
\begin{center}
{\LARGE On an asymptotic estimate of the $n$-loop correction in
perturbative QCD}\\[0.6in]
{\sc J.Ch\'{y}la, J.Fischer} and {\sc P.Kol\'{a}\v{r}}
\\[0.1in]
{\it Institute of Physics, Czechoslovak Academy of Sciences}\\
{\it Prague, Czechoslovakia}\footnote{Postal address: Na Slovance 2,
180~40~~Prague 8, Czechoslovakia}\\[0.4in]
{\bf Abstract}
\end{center}
\vspace{0.05in}
\noindent
A recently proposed method of estimating the asymptotic behaviour of
QCD perturbation theory coefficients is critically reviewed and shown
to contain numerous invalid mathematical operations and
unsubstantiated assumptions. We discuss in detail why
this procedure, based solely on renormalization group
(RG) considerations and analyticity constraints, cannot lead to such
estimates. We stress the importance of correct renormalization scheme
(RS) dependence of any meaningful asymptotic estimate and argue that
the unambiguous summation of
QCD perturbation expansions for physical quantities requires
information from outside of perturbation theory itself.
\newpage
\section{Introduction}
Because of computational complications only the lowest two or three terms
in QCD perturbation expansions are available for phenomenologically
interesting quantities. Taking into account that the relevant expansion
parameter (strong coupling constant $\alpha_{s}/\pi$) is of the order
of $0.1$,
and for certain quantities even bigger, we face the following two important
questions:
\begin{enumerate}
 \item how to reliably estimate higher order terms in perturbation expansions
   of physical quantities and, having obtained them,
 \item how to formulate a physically well-motivated summation method in the
    situation when these asymptotic estimates prevent the use of the familiar
Borel summation technique.
\end{enumerate}

A number of theorists have struggled over the last two
decades to evaluate multiloop Feynman diagrams in QCD and other theories and
we seem currently to be at the border of what can be done with the available
analytical as well as numerical tools. The state of the art in this field
has recently been reached by the first three loop QCD calculation
of a physical quantity, the familiar $R$-ratio in
 $\rm e^+\rm e^-$
annihilation into hadrons \cite{Andrei}
\begin{equation}
 R(q^{2})=\frac{\sigma({\rm e}^{+}{\rm e}^{-}\rightarrow
 {\rm {hadrons}})}
{\sigma({\rm e}^{+}{\rm e}^{-}
      \rightarrow\mu^{+}\mu^{-})}=\left(3\sum_{i=1}^{n_{f}}Q_{i}^{2}\right)
  \sum_{n=0}^{\infty}r_{n}(q^{2}/\mu^{2})\left(\frac{\alpha_{s}(\mu^{2})}
      {\pi}\right)^{n},
\label{R(Q)}
\end{equation}
where $q^{2}$ denotes the square of the center of mass energy in
${\rm e}^{+}{\rm e}^{-}$ collisions and $Q_{i}$ is the electric
charge of the produced quark with flavour $i$.
These, as well as the analogous results on certain structure functions sum
rules \cite{Larin} have been used in phenomenological analyses of
experimental data and the inclusion of NNLO correction has been shown
 to reduce theoretical uncertainties
\cite{e+e-,sumrule} and, in the case of the Gross-Llewelyn-Smith sum
rule, also to improve agreement with experiment \cite{sumrule}. There
seems, however, to
be little
chance of going  in a foreseeable future to still higher orders for
any of the physically measurable quantities. The quest for at least a
reliable estimate of these higher order contributions is therefore
highly commendable, in particular in view of the fact that the
relevant perturbation series are  expected to be factorially
divergent with asymptotically sign-definite terms.

In fact the question of the possible divergence of QCD (as well as other
field theories) perturbation expansions, which goes back to the original
argument of Dyson \cite{Dyson}, still defies definite and clear answer.
The original conjecture that the divergence of perturbation expansions is
directly related to the discontinuities of Green's functions in unphysical
domains of the coupling constant \cite{Parisi} has been seriously questioned
by several authors \cite{Khuri,Stevenson,Maxwell}. The knowledge of the
discontinuity itself is simply insufficient for the determination of
large order behaviour of perturbation theory coefficients and
one needs much more detailed information on the behaviour of Green's
functions in the vicinity of the corresponding cut in order to establish this
relation.

We share the viewpoint of Stevenson \cite{Stevenson}
that there is  little evidence that QCD perturbation
expansions for physical quantities do indeed diverge.
The examples discussed in \cite{Stevenson} indicate that in fact
most of the information
on the analytic properties of relevant Green's functions
near $g^2=0$ \cite{Hooft,Mueller} may actually be invisible in perturbation
expansions themselves.
Moreover, the fact, contained already in \cite{Khuri} and
later rediscovered and advocated by Stevenson \cite{Stevenson},
that the summability of perturbation expansions is not the same as the
possibility of recovering from such a series the corresponding ``full''
physical
quantity, has now gained wider acceptance. As the summability of QCD
perturbation expansions is determined by the large
order behaviour of their coefficients we primarily need to know which kind of
information is necessary and sufficient to derive it.

Quite recently two new attempts have been undertaken in order to derive the
asymptotic behaviour of the perturbation-theory coefficients for certain
physical
quantities \cite{Maxwell,West1}. They use completely different ideas but both
arrive at the same conclusion: QCD perturbation expansions are indeed
factorially divergent. The novel argument of \cite{Maxwell} is tailored
for the special case of particle fractions in ${\rm e}^{+}{\rm e}^{-}$
annihilation and
can not be used for such simple quantity as (\ref{R(Q)}). Contrary to
the original approach of \cite{Parisi} it, however,
doesn't employ any information on analytic properties of corresponding
Green's functions. Although it is based on several strong and questionable
assumptions, which render its conclusions open to doubt, we can at least
imagine that more sophisticated procedure can be
formulated along similar lines.

This is not the case with the procedure suggested in \cite{West1}, which is
claimed to answer both of the above questions.
In a subsequent paper
\cite{West2}, West uses his estimates to formulate a summation procedure with
truly remarkable properties. In particular they make
the use of a few lowest terms in (\ref{R(Q)}) a reliable approximation even
in the low $q^{2}$ region where the expansion parameter
$\alpha_{s}(q^{2})/\pi$ is so large that this approximation is normally
considered meaningless.

Soon after it appeared, \cite{West1} was criticized by Brown and Yaffe
\cite{Brown}, who have argued that the assumptions employed by West, i.e. the
analyticity constraints coupled with RG considerations, allow one to find
the relation between the Borel transform of (\ref{R(Q)}) and that of the
dispersive part of $\Pi(q^{2})$ but are not sufficient to unambiguously
determine the asymptotic behaviour of the  coefficients $r_{k}$
themselves. We
fully agree with these authors but consider it useful and illuminating to go
through the procedure of \cite{West1} carefully step by step in order to
see
exactly  {\em where\/} and {\em why\/} it breaks down if carried out
correctly. This will
be done in Section 3. Our analysis will also
show why there is no hope to remedy the derivation in \cite{West1}.
Finally in Section 4 we shall comment on the summation procedure suggested in
\cite{West2},
which uses the factorially divergent nature of the coefficents $r_{k}$ but is
in principle indepedent of the particular way of deriving it. We shall argue
that this algorithm is just one of an infinite number of equally plausible
``sums'' of (\ref{R(Q)}) compatible with the factorial growth of $r_{k}$.

In renormalized field theories like QCD the question of the large order
behaviour of the coefficients like $r_{k}$ in (\ref{R(Q)}) is
complicated by their inherent RS dependence.
Consequently even the convergence property of the expansion (\ref{R(Q)})
does in fact crucially depend on the RS chosen.
We therefore start our analysis with general remarks concerning the
constraints following from the formal RG invariance of expansions like
(\ref{R(Q)}) which any asymptotic estimate of $r_{k}$ must obey in order to
make sense.

\section{RG constraints and RS dependence of asymptotic estimates}
The expansion parameter $\alpha_{s}$ as well as the coefficients $r_{k}$ in
(\ref{R(Q)}) depend on the RS chosen. A part of this dependence is labelled by
the scale $\mu$ but the full RS ambiguity is much wider. As this point is
often a source of misunderstanding  let us specify very carefully what
is understood under this term and where it differs from the so called
{\em renormalization convention\/} (RC), the notion introduced in \cite{PMS}.
We have decided to include this detailed exposition after getting
acquainted
with West's response \cite{Westreply} to \cite{Brown} and other
critics \cite{Barclay}.

We start with the usual RG equation
\begin{equation}
 \frac{{\rm d}g(\mu^{2})}{{\rm d}\ln\mu} \equiv \beta (g)=
-g^{3}\left(b_{1}+b_{2}g^{2}+b_{3}g^{4}+\cdots\right),
\label{RG}
\end{equation}
which represents basically a definition of the expansion parameter
\begin{displaymath}
a(\mu^{2}) \equiv \alpha_{s}(\mu^{2})/\pi = g^{2}(\mu^{2})/4\pi^{2}
\end{displaymath}
used in (\ref{R(Q)}).
While $b_{1}$ as well as $b_{2}$ are unique numbers, fixed by the number
$n_{f}$ of quark flavours  (we consider massless quarks only), all
the higher order coefficients $b_{n}$, $n>2$, are arbitrary
numbers {\em defining\/} the renormalization convention.
For physical quantities the notion of the renormalization {\em scheme\/}
involves, besides the arbitrariness in the $\beta$-function coefficients,
another degree of freedom, not related to them. To exemplify this point
let us just recall the difference between the MS and
 $\overline{\rm {MS}}$
\ RS. Both RS have {\em exactly\/} the same $\beta$-function coefficients
but they are nevertheless associated with different values of the couplant
$a(\mu^{2})$ as well as the coefficients $r_{k}$,$k\geq2$.
In fact there is an infinite set of RS
which have the same $\beta$-function as
$\overline{\rm {MS}}$, but which correspond
to different values of $r_{k},k\geq2$. This dependence of the coefficients
$r_{k}$ on the RS chosen is formally to the k-th order
compensated by that of the couplant $a(\mu^{2},\rm {RS})$
in a way which guarantees the internal consistency of perturbation theory
\cite{PMS}. There are various ways of parametrizing this degree of freedom,
the most conventional one using just the value of renormalization scale
$\mu$.

Mathematically, this additional degree of freedom in the definition of the
RS is related to the simple fact that the RG equation (\ref{RG}) defining the
renormalized coupling constant $g(\mu^{2})$ (and thus $\alpha_{s}(\mu^{2})$)
has, even for fixed values of all the coefficients $b_{i}$, an infinite
number of solutions. For any
well-defined r.h.s. of (\ref{RG}), like for instance in the so called
't Hooft's RC, (or in any of the ``analytical'' RC discussed in
\cite{Barclay}) where all free higher order coefficients $b_{i},i\geq3$ are
set identical to zero \cite{Hooft}, the behaviour of $g(\mu^{2})$ for
$\mu\rightarrow\infty$ is essentially fixed by the first term in (\ref{RG}).
At finite $\mu$, however, the differences between various solutions of
(\ref{RG}) may be arbitrarily large.

Specifying the RS then means besides fixing the $\beta$-function coefficients
also choosing the renormalization scale $\mu$. Variation of
$\alpha_{s}(\mu^{2})$
with this scale is then compensated to any finite order by that of the
coefficients $r_{k}$. For a fixed $\mu$ there is, however, an infinite set of
the values $g(\mu^{2})$ corresponding to different
solutions of (\ref{RG}). So even setting $\mu$ equal to some external
kinematical variable (like the total CMS energy $\sqrt{q^{2}}$ in
(\ref{R(Q)})), we still face
the necessity to specify which of the solutions to (\ref{RG}) we have in
mind. This fact is frequently overlooked and one often finds statements that
the fixing of the scale $\mu$ and the $\beta$-function coefficients
$b_{k}$ defines uniquely the renormalization scheme. This is, however,
inaccurate and the selection of the ``referential'' RS, i.e. the selection of
one of the solutions to (\ref{RG}) as referential is as important as the
choice of the scale $\mu$ itself. Indeed, the variation of the scale $\mu$ for
fixed referential RS (RRS) is equivalent to the variation of the RRS for
fixed $\mu$. In the approach suggested and advocated in \cite{PMS}, we
select once for all one of these solutions and label different RS
by different values of $\mu$ and $b_{j}$. The selection of this referential
RS is of course  merely a matter of bookkeeping
and has nothing to do with the actual choice of a particular RS. The RS is
thus
uniquely defined by either of the sets \{$\mu,b_{i},\rm {RRS}$\},
\{$a,b_{i}$\} or \{$r_{2},b_{i}$\}. In the last two cases there is no need
to select the RRS.

There are various ways how to label the solutions of (\ref{RG}), the most
convenient
one employing the renormalization group invariant, dimensional parameter
denoted usually $\Lambda$. Each of the solutions of (\ref{RG}) is then
associated
with one particular value of $\Lambda$ (we denote it
 $\Lambda_{\rm {RRS}}$) and the
couplant $a(\mu^{2})$ is then
in fact a function of the ratio $\mu^{2}/\Lambda^{2}_{\rm {RRS}}$.
The precise definition of $\Lambda_{\rm {RRS}}$ being again a matter
 of convention, we can define it, for
example, by the relation $a(\mu^{2}=\Lambda^{2}_{\rm {RRS}})=\infty$.
Knowing the dependence of the couplant $a(\mu^{2}/\Lambda^{2};
\rm {RRS})$  on both the
scale and the RRS we come now to that of the coefficients $r_{k}$. For them
the internal consistency of perturbation theory implies the following
recurrence relations \cite{PMS}:
\begin{equation}
 \frac{{\rm d}r_{k}(q^{2}/\mu^{2};
 {\rm {RRS}})}{{\rm d}\ln\mu}=
 8\pi^{2}b_{1}(k-1)r_{k-1}
+32\pi^{4}b_{2}(k-2)r_{k-2}+{\cal O}(b_{3}).
\label{drk}
\end{equation}
For $r_{2}$ this yields (supplemented with the initial condition
$r_{0}=r_{1}=1$) the following explicit dependence on $\mu$
\begin{eqnarray}
\label{r2}
r_{2}(q^{2}/\mu^{2};{\rm {RRS}}) & =
 & 4\pi^{2}b_{1}\ln(\mu^{2}/q^{2})
+r_{2}(1;{\rm {RRS}}) \\
  & = & 8\pi^{2}b_{1}\ln(\mu/\Lambda_{{\rm {RRS}}}) -\rho_{1}(q^{2}),
   \nonumber
\end{eqnarray}
while for $r_{3}$ we get
\begin{equation}
r_{3}=\rho_{2}-c_{2}+(r_{2}+c/2)^{2},
\label{r3}
\end{equation}
where $c=b_{2}/b_{1}$ and $c_{2}=b_{3}/b_{1}$.
As a consequence of (\ref{drk}) analogous
algebraic relations hold also for
higher order coefficients $r_{k}$.
Similar consistency conditions can also be obtained
for the variation of $r_{k}$ with respect to other RS parameters, i.e.,
$b_{i}, i\geq3$.

The quantities $\rho_{1}$ and $\rho_{2}$ (and analogous quantities at
higher orders) entering the formulae (\ref{r2}) and (\ref{r3}) are RS
invariants,
$\rho_{1}$ being a function of $q^{2}$ while
$\rho_{2}$ (and all higher order invariants $\rho_{i},i>2$ as well)
is a pure number.
As $\mu$ and $r_2$ are (in a given RRS) in one-to-one correspondence
we can, instead of $\mu$, alternatively use the value of $r_2$ itself
to label the same degree of freedom as $\mu$. In other words, the
fixing of $r_2$ is an alternative way of fixing (together with the
coefficients $b_n$, $n>2$) the RS.
We stress that it is already the value of $r_{2}$ which is different in
the $\overline{\rm {MS}}$ and the ${\rm {MS}}$ RS, although
 the corresponding
$\beta$-functions have exactly the same coefficients $b_{i}$.

We mention this alternative labelling of the RS by means of \{$r_{2},b_{i}$\}
for the following reason.
It is in fact difficult to translate the results on the coefficients
$r_{k}$ obtained by means of any method not explicitly using Feynman diagram
renormalization (like the one in \cite{West1}) into the
language of the conventional perturbation theory and vice versa.
For instance, if, in West's approach, one were to carry out
the evaluation of $r_{k}$ in, say, $\overline{{\rm {MS}}}$ RS\
there seems to
be no obvious meaning of the term ``minimum subtraction'', which is inherent
to the usual Feynman rules calculations. The
same holds of course for any other conventional RS.
Using (\ref{r2}) we can, in any
selected RRS, define $\mu$ associated with the
 $\overline{\rm {MS}}$ RS by
demanding that the corresponding $r_{2}(\mu^{2}/q^{2};
{\rm {RRS}})=1.41$ but
this is possible only if the dependence of $r_{2}(\mu^{2}/q^{2};
{\rm {RRS}})$ on
$\mu$ is consistent with (\ref{drk}). In the notation where $\mu$ is fixed,
being set equal to, say, $\sqrt{q^2}$, similar statements hold for the
dependence on the RRS. In \cite{West1}, however, the asymptotic behaviour of
the coefficients $r_{k}$
\begin{equation}
r_{n}(1)\approx-\frac{e^{1+b'}}{\pi}(4\pi^{2}eb_{1})^{n-1}\;
       \frac{\Gamma(n+b')}{(n+b')^{2}},
\label{ras}
\end{equation}
is manifestly RRS independent.
Indeed it was the striking lack of any RRS dependence of (\ref{ras})
which has originally attracted our attention and cast doubts on their
validity. The absence of appropriate RRS dependence of $r_{n}(1)$, prevents,
however, the association of (\ref{ras}) with any well-defined RS. This crucial
fact has not been properly appreciated in \cite{West1}.

To see clearly the implications of the consistency conditions
(\ref{drk}) for the RRS dependence of the large order
behaviour of the coefficients $r_{k}$ let us consider
 the  (for technical reasons somewhat simplified) case when
$b_{k}=0$ for all $k>1$, i.e., also $b_{2}=0$. Then (\ref{drk}) can be
explicitly integrated to yield
\begin{equation}
r_{k}(q^{2}/\mu^{2};{\rm {RRS}})=
\sum_{j=0}^{k-1}\tau^{j}r_{k-j}(q^{2}/\mu_{0}^{2})\BC{k-1}{j},
\label{rsuma}
\end{equation}
where $\tau=8\pi^{2}b_{1}\ln(\mu/\mu_{0})$,
$\mu_{0}$ is some ``initial'' value of $\mu$ (corresponding to some
RRS$_{0}$)
and $r_{k}(q^{2}/\mu_{0}^{2})$ are the associated expansion coefficients of
(\ref{R(Q)}) in RRS$_{0}$. The above formula can simply be rewritten to
express the RRS dependence for fixed $\mu^2=q^2$:
\begin{equation}
r_{k}(1;{\rm {RRS}})=\sum_{j=0}^{k-1}\tau^{j}r_{k-j}
(1;{\rm {RRS}}_0)
\BC{k-1}{j},
\label{rsumab}
\end{equation}
where now
\begin{equation}
\label{tau}
\tau=8\pi^{2}b_{1}\ln(\Lambda_{{\rm {RRS}}_0}/
\Lambda_{{\rm {RRS}}}).
\end{equation}

Assuming, as an example, the factorial behaviour
of the coefficients, $r_{k}(1;{\rm {RRS}_{0}})=k!$,
 we get from (\ref{rsumab})
for general RRS
\begin{equation}
r_{k}(1;{\rm {RRS}})=k! \sum_{i=0}^{k-1}\frac{\tau^{i}}{i!}
\;\;\Rightarrow \;\;\lim_{k\to \infty}\frac{r_{k}}{k!}=\exp\tau.
\label{rtau}
\end{equation}
This means that although the character of the large order behaviour of
the coefficients $r_{k}(1;{\rm {RRS}})$ is not changed (in the above
example it is merely multiplied by the $k$-independent factor $\exp\tau$)
when we
vary the RRS, the magnitude of the coefficients
$r_{k}(1;{\rm {RRS}})$ obviously
is. As $\tau$ can in principle be any real number, the magnitude of the
coefficients $r_{k}(1;{\rm {RRS}})$ can, in our example,
 be arbitrarily large.
Although the
RRS dependence of the large order behaviour of $r_{k}(1)$ is irrelevant
for considerations concerning the summability of our series, it is
crucial when we are interested in their numerical values. For the realistic
case with $b_{2}\neq 0$ the dependence of
$r_{k}(1;{\rm {RRS}})$ on the RRS is
more complicated than (\ref{rtau}) but its main feature i.e. the
arbitrariness of the magnitude of $r_{k}(1;{\rm {RRS}})$
on the RRS persists.
We want to  emphasize this point very strongly as in his reply
\cite{Westreply} West explicitly claims that the large order behaviour of the
coefficients $r_{k}(1)$ should be ``scheme invariant'' with ``scheme
dependence'' entering only via the dependence of nonleading corrections on
the free coefficients $b_{k},k\geq 3$. The relation (\ref{rtau}) demonstrates
that this claim is definitely wrong as the dependence
 of $r_{k}(1;{\rm {RRS}})$
on the RRS may change the magnitude of them essentially at will. For
example,
 the value of $\tau$ corresponding to the change from the
$\overline{\rm {MS}}$ to the MS RRS equals, for $n_f=5$
\begin{equation}
\label{taunum}
\tau=8\pi^{2}b_{1}\ln(\Lambda_{\overline{\rm {MS}}}/
\Lambda_{{\rm {MS}}})
=3.83 \ln 2.65 = 3.73.
\end{equation}

 There is thus no obvious reason to compare $r_{3}=-13.4$ from (\ref{ras})
with its exact value in the $\overline{\rm {MS}}$ RS (where
$r_{3}(1;\overline{{\rm {MS}}})=-12.8$) and not,
for instance, in MS which has exactly the same $\beta$-function coefficients
but where, however, $r_{3}(1;{\rm {MS}})=16.3$.
 The agreement of (\ref{ras}) with
$r_{3}(1;\overline{\rm {MS}})$ should therefore be
considered as a pure
coincidence.
This has also been demonstrated in \cite{Barclay} on the basis of the
comparison of the corresponding $n_{f}$ dependences of (\ref{ras})
 and the exact results \cite{e+e-}.

There is in fact no agreement of (\ref{ras}) even in the
 $\overline{\rm {MS}}$
RS and even following uncritically the whole derivation of \cite{West1}. The
``correct'' (i.e. disregarding all the objections we shall discuss in the
next section) result thereof reads
\begin{equation}
r_{n}(1)\approx-\frac{e^{1+b'}}{2\pi}(4\pi^{2}eb_{1})^{n-1}\;
       \frac{\Gamma(n+b')}{(n+b')^{2+b'}},
\label{rascor}
\end{equation}
which differs from (\ref{ras}) asymptotically by a factor $(n^{-b'})/2$
and by  a factor of about $1/4$ for $n=3$.
So even within the framework of West's approach the starting point
for his considerations, i.e., the coincidence with exact results of
\cite{e+e-}
is actually absent. On the other hand it is clear that for {\em any\/}
value $\tilde{r}_{3}$ there is always such a RRS that
$r_{3}(1;{\rm {RRS}})=\tilde{r}_{3}$. We also stress that
the ambiguity in association of (\ref{ras}) with a particular RS has nothing
to do with the neglected higher order $\beta$-function coefficients $b_{k}$
but reflects the fact that even for a given RC, i.e. for  fixed $\mu$ and
$b_{k},k\geq3$ there is still a degree of freedom connected with RRS.

In perturbation theory the quantity (\ref{R(Q)}) is described by a
{\em pair\/} of expansions (\ref{R(Q)}) and (\ref{RG}), neither of
which has a
physical meaning of its own. Except for the first two coefficients in
both (\ref{R(Q)}) and (\ref{RG}) all other ones are arbitrary,
subject only to the internal consistency constraints like (\ref{r3}) which
follow from (\ref{drk}).
We can choose the RS=\{$\mu,b_{i}$,RRS\} in such a way that either
(\ref{R(Q)}) or
(\ref{RG}) is a well-defined, convergent series but we can not do this for
both
of them simultaneously. For instance, the ``effective charges'' approach
of Grunberg \cite{Grunberg} sets $r_{k}=0$ for all $k>1$ so that one has to
worry about the convergence property of (\ref{RG}) only. On the contrary, in
't Hooft's RC \  $b_{k}=0,\ k>2$ so that the definition equation (\ref{RG})
of the couplant $a(\mu^{2})$ is well-defined  and
the nontrivial question concerns the expansion coefficients $r_{k}$. In a
computationally simple RS, like $\overline{\rm {MS}}$, both of the
 expansions (\ref{R(Q)}) and (\ref{RG})
are presumably divergent. This makes their use for the purpose of summation
considerations rather unsuitable as the couplant $\alpha_{s}/\pi$
itself is ill-defined in the limit of infinite order. Consequently, any
statement concerning the large order behaviour of the coefficients $r_{k}$
is inextricably connected with that of the coefficients $b_{k}$ of (\ref{RG})
and makes sense only when it is clearly and unambiguously related to some
definite RS.

\section{Discussion of West's approach}
We now turn to a discussion of the West approach advocated in \cite{West1}.
(For reader's convenience we basically use the original notation of
\cite{West1}.)
We start from the definition of the vacuum polarization function
$\Pi(q^2,g^2)$ in terms of the time-ordered product of currents:
\begin{equation}
\label{bas}
(q_{\mu}q_{\nu}-g_{\mu\nu} q^2)\;\Pi(q^2,g^2)=i\int dx
\; e^{iqx}\;\Vev{Tj_\mu(x)j_\nu(0)},
\end{equation}
where $j_\mu(x)$ is the electromagnetic current of the quarks in massless
QCD. The function $\Pi $ is assumed to be analytic in the whole $q^2$
complex plane cut along the positive real axis.
 The
imaginary part of $\Pi$ is related to the ratio $R$ (see eq.\ (\ref{R(Q)}))
\begin{equation}
\IM\Pi(q^2/\mu^2,\alpha_s(\mu^2)) =
\frac{1}{12\pi}R(q^2/\mu^2,\alpha_s(\mu^2)).
\end{equation}
 We further
introduce the function
\begin{displaymath}
    D(q^2/\mu^2,g^2)=\frac{\partial}{ \partial t}\; \Pi(q^2/\mu^2,\alpha_s ),
\end{displaymath}
where $t=\ln(q^2/\mu^2)$.
$R$ and $D$ satisfy the homogeneous renormalization group
equation
\begin{equation}
    [\mu \frac{\partial}{\partial \mu }+\beta (g)
\frac{\partial}{\partial g }]\; R(q^2/\mu^2,\alpha_s) = 0
\end{equation}
and analogously for $D$,
where the Callan-Symanzik function $\beta$ is defined in (\ref{RG}).
As a consequence
the functions $R$ and $D$
depend  (in massless QCD) on the single variable
$z=(q^2/\mu^2)e^{2K(g)}$ \cite{Hooft,West1}. Here the function $K(g)$ is
defined by

\begin{equation}
\label{k}
    K(g)=\int_{g_0}^{g}\,\frac{dg'}{\beta(g')}\approx \frac{1}{2 b_1
g^2} - \frac{b'}{2}\ln(b_1/g^2 + b_2)+\,\cdots ,
\end{equation}
where $b'=b_2/b_1^2$. (In obtaining eq.\ (\ref{k}) a convenient choice of
$g_0$ has been made.) It is useful to define the running coupling
constant $\gr$
 by $2K(\gr)-2K(g)=t$. Then the eq.\ (\ref{R(Q)}) can be rewritten as
\begin{equation}
\label{Rexp}
R(q^2/\mu^2,\alpha_s(\mu^2))=R(1,\al)\approx (3\sum_i
Q_i^2)\sum_{n=0}^{\infty}\, r_n (1)\,(\al /\pi)^n.
\end{equation}

The dispersion relation for $D$ can be given the form
\begin{equation}
\label{disp}
    D(q^2/\mu^2,g^2)=A(q^2/\mu^2,g^2)\int_0^\infty
dz\,\frac{f(z)}{[z-A(q^2/\mu^2,g^2)]^2},
\end{equation}
where
\begin{equation}
    f(z)=\frac{1}{12\pi^2}\; R(q^2/\mu^2,\alpha_s(\mu^2)),
\end{equation}
and
\begin{equation}
    A(q^2/\mu^2,g^2)=\frac{q^2}{\mu^2}\,e^{2K(g)}.
\end{equation}
(In the following we shall use the reduced notation $A(k)=
A(q^2/\mu^2,1/k)$, where $k=1/g^2$.)

The aim of West's considerations is to use eq.\ (\ref{disp}) in the analysis
of large $n$ behaviour of the expansion coefficients $d_n$ defined by formal
(asymptotic) expansion
\begin{equation}
\label{asexp}
D(q^2/\mu^2,g^2)\approx\sum_{n=0}^{\infty}\,(-1)^n d_n (q^2/\mu^2)\,g^{2n}.
\end{equation}
The expansion coefficients $r_n$, defined in (\ref{R(Q)}), are then
connected with the  $d_n$ by the relation
\begin{equation}
\label{r:d}
\left( \sum_{i}Q_i^2\right)\, r_n=-(-4\pi ^2)^{n+1}\frac{1}{\pi n
b_1} [\IM
d_{n+1}+\frac{b_2}{b_1}\;\IM d_n +\cdots],
\end{equation}
as a consequence of the equation
\begin{equation}
\label{ImD:R}
    \IM D = \frac{g\beta (g)}{12\pi }\;\frac{\partial R}{\partial g^2}
\approx -(\sum_{i} Q_i^2)\;\frac{b_1 g^4}{16\pi^3}
[1+(\frac{b_2}{b_1}+\frac{r_2}{2\pi^2})g^2 +\cdots ].
\end{equation}
This suggests that
 we can  get the large order behaviour of $r_n$
 once we know that of $d_n$ (see however \cite{Brown}).

We now focus on three crucial points
in the method of calculation of the coefficients $d_n$.

{\bf(A)} If eq.\ (\ref{asexp}) can be understood as an
asymptotic expansion of $D$ then the $d_n$ are given by (see Wightman
\cite{Hooft})
\begin{equation}
\label{dcoeff}
    d_n=-\lim_{g_0\to 0^+} \oint_{C(g_0)}\frac{dg^2}{2\pi
i}(-g^2+g^2_0)^{-1-n}\;D(q^2/\mu^2,g^2).
\end{equation}
$C(g_0)$ is a closed contour around the point
 $g_0^2$.
The size of the contour has to be chosen sufficiently small in order not
to hit the complicated singularity structure  of the function $D$ around
the origin. In particular, $C(g_0)$ must not leave, with $g_0$
approaching zero, the wedge bisected by the real axis and bounded above
and below by circles that are tangent to the real axis at the
origin\cite{Hooft}.
(There are arguments that this horn-shaped region is the analyticity
region of the function $D$ \cite{Hooft}.)
Making the transformation $k=1/g^2$
and choosing the corresponding contour $C'(k_0)$ around the point
$k_0=1/g_0^2$
(see Fig.1),
we can rewrite eq.\ (\ref{dcoeff})
into the form
\begin{equation}
\label{dcoeff:k}
    d_n=\lim_{k_0\to \infty} \oint_{C'(k_0)}\frac{dk}{2\pi
i}\; k^{n-1}\; k_0^{n+1}\; (k-k_0)^{-1-n}\;D(q^2/\mu^2,1/k).
\end{equation}
One particularly convenient choice of $C'(k_0)$ is the contour $C'_0$
(see Fig.1) which is independent of the position of $k_0$
provided that $k_0$ is sufficiently large.
 (We denote as
$C_0$ the corresponding contour in the $g^2$--plane.)
We do not
 extend the contour $C_0'$ up to $\RE k\to -\infty$, in order to
 avoid the problem with a singularity of $D$ at $|k|=\infty $. Thus,
in  the contour $C_1'$ depicted in Fig.1
 one cannot  neglect the contribution
along the direction perpendicular to the real axis at $\RE k\to -\infty$.
The contour mentioned in
 \cite{Westreply} should therefore be understood as the
 $\RE k\to -\infty$ limit of $C_1'$.

 Obviously in eqs.(\ref{dcoeff}) and
(\ref{dcoeff:k}) one can not interchange the order of the limit
$k_0\to\infty $ and
the integration. Otherwise we would obtain the following result
\begin{equation}
\label{westint}
    d_n=\oint_{C_0'}\frac{dk}{2\pi
i}(-k)^{n-1}\;D(q^2/\mu^2,1/k).
\end{equation}
Using the expansion (\ref{asexp}) in the neighbourhood of $\RE k\to
-\infty$ we conclude that the integral (\ref{westint}) does not exist
for any $n\ge 0$.
Rewriting (\ref{westint})  back to the $g^2$--plane
we obtain the expression
\begin{equation}
\label{westintg}
    d_n=-\oint_{C_0}\frac{dg^2}{2\pi
i}\; (-g^2)^{-n-1}\;D(q^2/\mu^2,g^2),
\end{equation}
which coincides with  West's formula (with the integration contour
properly defined). Thus, the $d_n$ presented in
\cite{West1} cannot be understood as well-defined quantities.

{\bf (B)}
The next step in West's approach is the substitution of (\ref{disp}) into
(\ref{dcoeff:k})
\begin{equation}
\label{dn}
    d_n=\lim_{k_0\to \infty} \oint_{C_0'}\frac{dk}{2\pi
i}k^{n-1}k_0^{n+1}(k-k_0)^{-1-n}\; A(k)\int_0^\infty
dz\,\frac{f(z)}{[z-A(k)]^2},
\end{equation}
and the interchange of the order of integrations. We shall however show that
the $d_n$
computed in this way are incorrect (this suspicion was also voiced in
ref.\cite{Barclay}).
For this purpose we introduce the
quantity
\begin{equation}
\label{dbar}
    \dnn = \lim_{k_0\to\infty}\;\int_{0}^{\infty} dz\;
f(z)\; \frac{\partial}{\partial t}\;\Phi(n,t,z,k_0),
\end{equation}
where
\begin{equation}
\label{newphi}
\Phi(n,t,z,k_0) =\frac{1}{2\pi i}\oint_{C_0'} dk\;
\frac{k^{n-1}\;k_0^{n+1}}{(k-k_0)^{n+1}}\; \frac{1}{z-A(k)}.
\end{equation}
Note that the formal limit $k_0\to\infty$ would recast this relation to
the form
\begin{equation}
\label{westphi}
\Phi(n,t,z) =\frac{1}{2\pi i}\oint_{C_0'} dk\;
\frac{(-k)^{n-1}}{z-A(k)},
\end{equation}
used in \cite{West1}.

Applying the Cauchy theorem to (\ref{newphi}) we get
\begin{eqnarray}
\label{twophi}
\Phi(n,t,z,k_0)&=&
k_0^{n+1} \left\{ \frac{1}{n!}\; \frac{\partial^n}{\partial k^n}\;
\frac{k^{n-1}}{z-A(k)}\right|_{k=k_0}- \nonumber  \\[2ex]
 &-& \left. \frac{k_p^{n-2}(b_1 k_p  + b_2)}{(k_p - k_0)^{n+1}A(k_p)}
\;\delta_{C'} \right\} \\[2ex]
 & \equiv & \Phi_1 (n,t,z,k_0) + \Phi_2 (n,t,z,k_0).\nonumber
\end{eqnarray}

The contour is chosen such that at most one of the poles given by
\begin{equation}
\label{pole}
A(k)=z-e^{t+2K(k)}=0,
\end{equation}
can lie inside it. We will denote its position by $k_p$
(see Appendix).
$\delta _{C'}$ is a contour dependent factor, $\delta _{C'}=1$ if the pole
is inside $C'$, otherwise $\delta _{C'}=0$. This contour-dependence is
apparently undesirable and it is the first signal that $\dnn$ given by
(\ref{dbar}) differs from $d_n$ ($d_n$ is contour-independent
provided that $k_0$ lies inside $C'_0$).

Using (\ref{twophi}) it is easy to show by induction over $n$ that
\begin{equation}
    \lim_{k_0\to\infty}\int_{0}^{\infty}dz\; f(z) \frac{\partial}{\partial t}
\;
\Phi_1 (n,t,z,k_0)=d_n.
\end{equation}
Thus, we get
\begin{equation}
\label{diffd}
\dnn - d_n = \lim_{k_0\to\infty}\int_{0}^{\infty}dz\; f(z)\;
\frac{\partial}{\partial t} \;
\Phi_2 (n,t,z,k_0).
\end{equation}
We conclude that this contour-dependent difference is given by the pole
contribution to $\dnn$. The position of the pole
depends on the value of $z$. (When $b_2\approx 0$ we have
$k_p\approx b_1(\ln z-t)$.) In general, the pole moves to the right in the
$k$--plane when
$z$ increases.
 The integration in (\ref{diffd}) can therefore be restricted to an interval
$(z_1,z_2)$  defined by the occurrence of pole inside $C'$. In
the case of the contour $C'_0$ depicted in Fig.1 we have $z_2=\infty $.
For our aims
it is sufficient to prove that the difference (\ref{diffd}) does not
vanish.
 In fact we prove
in the appendix that the difference can be  divergent for
some contours. Consequently, one cannot interchange
the order of integrations in (\ref{dn}). The same objection has been
raised
in \cite{Barclay} (in this paper, however, the contribution of the pole to
$\dnn$ is  ignored).

{\bf (C)}
In \cite{West1}
the integral (\ref{westphi}) has been estimated by the
saddle point technique which  might be expected to
 provide a large-$n$ approximation  to (\ref{twophi}).
An essential feature within the saddle point method is a convenient choice of
the integration contour.
Representing the integrand in the
form of $\exp h(k)$, where $h(k)$ is a complex function analytic in some
vicinity of the contour, we should choose the contour so that
the following
conditions are satisfied:
\begin{description}
    \item[{\rm (i)}] The integration contour can be deformed (without changing
the value of the integral) so that it passes through the
saddle point $k_s$.
    \item[{\rm (ii)}] $\RE h(k)$ has its maximum on the
contour just at the saddle point $k_s$.
    \item[{\rm (iii)}] The contour can not be deformed so that a new maximum
of $\RE
h(k)$ is below $\RE h(k_s)$.
\end{description}

A detailed inspection of the integrand in
(\ref{westphi})
reveals the fact that the
condition (ii) is not satisfied if we integrate in the direction
perpendicular to the real axis in the complex $k$--plane. The curve of the
steepest descent passes along the real axis. This  fact
invalidates the approach of \cite{West1} where the
integration contour is not specified and seems to be chosen
perpendicular to the real axis at the
saddle point. Moreover,
a detailed analysis of the saddle points shows their strong dependence
on the value of $z$. This crucial fact was also overlooked in
\cite{West1}, where $k_s\approx b_1(n-1+b')$ was taken. We
illustrate the situation in Figs.2-5, where the positions of poles and
saddle points of the integrand in (\ref{westphi}) are displayed for
different
values of $z$ and for $n=4$. One can notice
an interesting
interplay of poles and saddle points
resulting in the situation when two complex
conjugate saddle points become leading ({\em i.e.}\ with maximal
$\RE k$) for sufficiently large $z$. The
$z$ dependence of the saddle point $k_s=k_s(z)$ does not allow one
 to choose  any universal ($z$-independent) integration contour in the
$k$--plane. Hence,
the use of
the saddle point method is plagued with the occurrence of the
moving leading saddle point $k_s(z)$. Consequently, one cannot obtain
the relation
($\Phi (k)$ is defined
in \cite{West1})
\begin{equation}
\label{west:cruc}
d_n(q^2/\mu^2)\simeq \frac{1}{2} \left[ \frac{2}{\pi \Phi (k_s)}\right]^{1/2}
k_s^{n-1} D(q^2/\mu ^2,1/k_s) \cos n\pi ,
\end{equation}
which plays a crucial role in the derivation of West's result. (We have
added the factor $1/2$ which is missing in \cite{West1}.)
It was just this relation
which made it possible, employing (\ref{r:d}) and (\ref{ImD:R}), to
derive the simple relation (\ref{ras})
for the coefficients of the expansion (\ref{Rexp}). We would like to
stress that even within West's approach, based on the saddle point
method, the $d_n$ will also get
contributions that cannot simply be related to the value of
$D(q^2/\mu^2,1/k)$ for some large $k$ (in (\ref{west:cruc}) $k_s\to\infty$
when $n\to\infty $). This follows from the $z$-dependence of the leading
saddle point $k_s$ mentioned above.
 The same argument
can also be  used for the case of finite $k_0$ in (\ref{newphi}).

Moreover, the pole of the integrand in (\ref{newphi}) generates, at
finite $k=k_0$, an additional saddle point which disappears when
$k_0\to\infty$, and for some $z$, becomes the leading one. For
these values of $z$ saddle point method can be used to estimate the
integral (\ref{newphi}) because the conditions (i)--(ii) are
satisfied.
This is another signal that the saddle point method of approximating
the $d_n$ is not applicable in the form advocated by West.

\section{Summing perturbation series in QCD}
Using the asymptotic estimate (\ref{ras}) West has in his subsequent paper
suggested \cite{West2} a
way of handling the divergence of QCD perturbation series which has two
remarkable properties:
\begin{enumerate}
\item it is extraordinarily close to the ``correct'' sum using only a few
 lowest order terms and, moreover,
\item the number of terms to be taken into account in his procedure is
inversely proportional to the value of the couplant, which implies that it
decreases as we approach the infrared region, i.e. as $q^{2}\rightarrow0$ in
(2).
\end{enumerate}
These properties are remarkable and, were they true, would have
serious repercussions as the conventional wisdom tells us that at low
$q^{2}$ the
contribution of nonperturbative effects
 are expected to become dominant \cite{Fatejev,Zacharov}. As the
message of \cite{West2} is, from the viewpoint of phenomenological
applications, very appealing we wish to point out several simple and
obvious facts which show the arbitrariness of West's construction.
Basically what he does in \cite{West2} is that he inadmissibly generalizes
the conclusions valid for one particular definition of the sum, which really
has the above-mentioned properties, to the case of QCD perturbation expansion
(\ref{R(Q)}). Let us recall the basic fact that for the divergent
asymptotic series
\begin{equation}
\sum_{n=0}^{\infty}a_{n}z^{n}
\label{aseries}
\end{equation}
there is no ``natural'', unique, sum. There is an infinite number of
functions
$F(z)$ which have the series (\ref{aseries}) as its asymptotic one and these
functions while sharing the same behaviour in the vicinity of $z=0$ may for
any fixed $z\neq 0$ be {\em arbitrarily\/} far apart. Perturbation theory
itself does not tell us which one of them to choose and we have to look
somewhere else for ideas what to do with it \cite{Zacharov,Aachen}.
Without first defining what we mean under the ``sum'' of the
asymptotic series (\ref{aseries}), there is no sense in talking about the
``error'' of
any finite order approximation. To claim, as is done in \cite{West2},
that this error, defined as
\begin{equation}
R^{N}=\left|F(z)-\sum_{n=0}^{N-1}a_{n}z^{n}\right|,
\label{error}
\end{equation}
can be estimated by
\begin{equation}
R^{N}\approx a_{N}z^{N}
\end{equation}
is therefore unsubstantiated.

 The asymptoticity of the series
(\ref{aseries}) merely means that
\begin{equation}
 \lim_{z\rightarrow0}{\frac{R^{N}(z)}{a_{N-1}z^{N-1}}}=0
\end{equation}
but, for a fixed $z\neq 0$, $R^{N}$ may be quite arbitrary. In
other words any statement concerning the error estimate (\ref{error}) is
meaningless unless we unambiguously specify the definition of the sum $F(z)$
appearing in (\ref{error}) as well. This holds equally well
for Borel summable as well as non Borel summable series. In \cite{West2} the
author bases his claims essentially on the property of a particular Borel
summable series with the coefficients $r_{k}=(-1)^{k}k!$. He demonstrates
that by choosing the number $N(a)$ of terms in (\ref{error}), where this
value is roughly inversely proportional to the expansion parameter $a$ and
given by the local minimum of the contribution of $a_{n}z^{n}$ to the partial
sum, he gets a very good approximation of the corresponding Borel sum.
This is by no means the only possible approach, even for Borel
summable series: one needs further information
on  the function $F(z)$ itself (i.e. information not contained
in the perturbation expansion) to guarantee the uniqueness of the
Borel sum. To generalize the properties
of this toy example series (coupled with one ad hoc definition
of the sum) to the non Borel summable series in realistic QCD is an entirely
arbitrary procedure. Moreover, the procedure suggested in \cite{West2} has
numerous concrete shortcomings. We mention just the most important ones:
\begin{itemize}
\item  It is obvious that by adding to this toy example series
any convergent one we get the expansion with the coefficients which have
the same large-$n$ behaviour while to any finite order the partial sums may
be arbitrarily far apart. Any meaningful statement concerning the accuracy of
a given summation technique must take into account the true values of the
expansion coefficients, not merely their asymptotic behaviour. To determine
the value of $N(a)$ from this asymptotic behaviour is unsubstantiated.
\item The simple picture in which the contributions of the terms $a_{n}z^{n}$
first monotonously decrease and then starting at $N(a)$ increase forever
is a property of the  toy example mentioned but in realistic QCD the behaviour
of this contribution may be much more complicated. In fact, for a function
$F(z)$ to have (\ref{aseries}) as its asymptotic series we can take
$N(a)$ entirely arbitrary, subject only to the condition that as
$a\rightarrow0$,\ $N(a)\rightarrow\infty$. For $a\to 0$ all these
procedures do, of course, converge to the same result but for any
{\em nonvanishing\/} $a$ they may yield vastly different numbers.
Consequently, also the estimate
(17) in \cite{West2} of the error of the  procedure suggested is valid for
the  example discussed but has no general validity.
To give his claims some justification West would have to argue why his
definition of the estimate is to be preferred to the infinite number
of other possibilities.
\item For the asymptotic series of the type $r_{k}\approx k!$,
 expected to
occur in QCD, there is no ``natural'' sum like the Borel one for the sign
alternating series $r_{k} \approx (-1)^{k}k!$. To define in these
circumstances the
estimate by means of the same, above mentioned, procedure is
unsubstantiated and has no physical motivation behind it.
\item Even if all the coefficients $r_{k}$ in (2) were known exactly (as in
the toy example of the series with coefficients $r_{k}=(-1)^{k}k!$), the
summation procedure suggested in \cite{West2} would still be itself badly
ambiguous as the value of $N(a)$ and consequently also of the estimate defined
as
\begin{equation}
 \sum_{k=0}^{N(a)} r_{k}(\alpha_{s}/\pi)^{k}
\end{equation}
would depend on the RS in that the truncation procedure is carried out.
\end{itemize}

\section{Summary and conclusion}
We have examined in detail the method recently proposed by West for
the determination  of the  large order behaviour
of perturbation expansion coefficients in QCD. Going carefully through
his derivation, which is based exclusively on the renormalization
group considerations and analyticity properties of the vacuum
polarization function, we have identified several points where strong
and largely unsubstantiated assumptions have to be made in order to
arrive at the result (\ref{ras});
in particular
\begin{description}
\item[{\rm (A)}] the interchange of the limit $k_0\to\infty $ and the
integration over $k$ in (\ref{dcoeff:k});
\item[{\rm (B)}] the interchange of integrations in the double integral in
(\ref{dn}), which results in the inadmissible contour dependence of  the
result;
\item[{\rm (C)}] the use of the saddle-point technique for  the  estimate
of the contour integrals,  in  spite  of the  ignorance  of  the
$z$-dependence of the positions of the leading saddle points.
\end{description}

 A first signal that there is something wrong with the result (\ref{ras})
 comes already from the fact that this formula for $r_n$ is
 independent of the RRS chosen. However, as is obvious from his reply
 \cite{Westreply} to the early critical voices \cite{Brown,Barclay},
West considers this
property of his formula (\ref{ras}) on the contrary a merit. To clarify this
crucial point we have discussed at length in particular the correct
RS dependence of any consistent large order behaviour of the
coefficients $r_k$.
In general, we thus agree with the claim of Brown and Yaffe\cite{Brown}
that the
information exploited by West is in principle insufficient for the
derivation of the asymptotic estimates of these coefficients.
Compared with their work, which shows what can be derived from these
assumptions, we have attempted to demonstrate why the construction
suggested by West doesn't in fact work if carried out with proper
care, and why there is no conceivable way to improve it and still
arrive at an asymptotic estimate of the type (\ref{ras}).
The large order behaviour of QCD perturbation expansions thus remains
to large extent an open question which certainly deserves further
attention. In our view, to give these expansions good mathematical as
well as physical meaning we have to go beyond the purely perturbative
framework and look, as in  \cite{Zacharov,Aachen,Chyla}, to their
interplay with various nonperturbative effects.

\vspace {0.6cm}
\section*{Acknowledgement}
We wish to thank A.Kataev for careful reading of the manuscript and
many useful comments and suggestions.

\appendix
\section*{Appendix}
First, let us derive the equation for the boundary of the principal strip
in the $k$-plane.
The relation
\begin{equation}
\label{transf}
q^2/\mu^2\;e^{2K(g)}=z
\end{equation}
transforms, at $q^2$ fixed, the complex $z$-plane into a strip in the
$k$-plane. $K(g)$ is given by (\ref{k}). Let us introduce the notation
\begin{equation}
 k= u+iv, \hspace{2em} q^2/\mu^2=\sigma e^{i\varphi},\hspace{2em}
 \ln \sigma=t,\hspace{2em}
z=re^{i\psi}
\end{equation}
and take the logarithm of (\ref{transf}) ($\ln$ stands for the
principal branch):
\begin{equation}
\ln\sigma + u/b_1 -\frac{1}{2}b' \ln[(b_1 u + b_2)^2 +b_1^2 v^2]=\ln r
\end{equation}
and
\begin{equation}
2\pi N +\varphi -\psi +v/b_1 - b' arg(b_1 k + b_2)= 0
\end{equation}

The borders of the strip are given by the inequality
\begin{equation}
-\pi \le \psi < \pi ,
\end{equation}
which represents the whole complex cut $z$-plane. Choosing $N=0$ and
$\varphi =0$, we obtain the implicit equation for the upper bound of
the strip in the form
\begin{equation}
v-\frac{b_2}{b_1}\arctan(\frac{b_1 v}{b_1 u+b_2}) - \pi b_1
-\frac{b_2}{b_1}\pi \theta (-b_1 u -b_2)=0,\ v>0,
\end{equation}
the lower border being symmetric with respect to the real axis.

We now turn to the contour dependence of the difference
defined in (\ref{diffd}):
\begin{equation}
\label{app:diffd}
\Delta d_n\equiv \dnn - d_n = \lim_{k_0\to\infty}\int_{z_1}^{z_2}dz\; f(z)\;
\frac{\partial}{\partial t} \;
\Phi_2 (n,t,z,k_0),
\end{equation}
where
\begin{equation}
\label{app:phi2}
\Phi_2 (n,t,z,k_0)=-k_0^{n+1}\;
\frac{k_p^{n-2}(b_1 k_p  + b_2)}{(k_p - k_0)^{n+1}A(k_p)},
\end{equation}
and  for $z\in (z_1,z_2)$ the pole $k_p$ occurs inside the integration
contour.
We note that a usual regularization is understood in the
denominator factor
$(k_p-k_0)$ of (\ref{app:phi2}): we can add a small imaginary part to
$k_0$.
The function $k_p(z)$ solves the equation (\ref{pole}).
Denoting $k_p=u_p+iv_p$ one
can show that the positions of poles
 satisfy the equation
\begin{equation}
\label{poleeq}
\frac{\vtil_p}{\tan (\vtil_p + \gamma) }-\frac{1}{2}\, \ln \vtil_p^2 +
\frac{1}{2} \ln [\sin^2(\vtil_p +\gamma )]-B=0,
\end{equation}
where
\begin{eqnarray*}
    \vtil_p & =& \frac{b_1}{b_2}\, v_p,\\[2ex]
 \gamma & = & 2\pi N b_1^2/b_2,\hspace{2em} N=0,\ \pm 1,\ \pm 2,\dots ,\\[2ex]
    B & = & 1+\ln b_2 + \frac{b_1^2}{b_2}\,(\ln z - t),
\end{eqnarray*}
provided that $\vtil_p +\gamma \neq n\pi ,\ n=0,\ \pm 1,\ \pm 2 \dots$.
Knowing $v_p$ the real part $u_p$ can then be calculated
from the equation
\begin{equation}
    \tilde{u}_p\equiv u_p\frac{b_1}{b_2} = \frac{\vtil_p}{\tan(\vtil_p +
\gamma) } - 1
\end{equation}
A different regime can be observed  if $\vtil_p
+\gamma = n\pi$. These are the relevant poles because the integration
contour is situated in the right half plane. (We consider
only such contours which can be
hit only by one pole, {\em i.e.\/}, which are placed in the strip.)
The poles  on the real axis
satisfy the equation
\begin{equation}
\label{realpole}
\tilde{u}_p-\frac{1}{2}\,\ln [(1+\tilde{u}_p)^2]-B=0
\end{equation}
Here the corresponding $v_p$ equals, of course, zero.
In the 5-flavour case, i.e., $b_1=0.0485$ and
$b_2=0.00155$, we get that for $z>71.024$ the equation (\ref{realpole})
has two solutions, one with $u_p$ positive, the other one with $u_p$ negative.
They move to opposite  sides if $z$ increases starting from the
origin. If $z<71.024$ we get no real poles in the  strip.
However, one can find two complex conjugate poles in the left half of the
complex $k$-plane. The $z$-evolution of the poles is depicted in
Figs.2-5. As is seen there are two poles which collide at
$z=71.024$ from either side along the imaginary axis and scatter
along the real axis.
In the following we will consider the pole moving along the
positive real axis; let us denote its position  by
$k_p(z)$, for $z>71.024$.

Taking the derivative of
(\ref{app:phi2}),
substituting the variable $y=k_p(z)$ into the integral over $z$ and defining
\begin{equation}
    F_{n,m}(k_0,y_1,y_2)=\int _{y_1}^{y_2}dy\;
\tilde{f}(y)\frac{y^n}{(y-k_0)^m}
\end{equation}
we get
\begin{eqnarray}
\label{delt}
    \Delta d_n &=&\lim_{k_0\to\infty}k_0^{n+1}\,\{ b_1^2 (n+1)
F_{n-1,n+2}+b_1 b_2
(n+1) F_{n-2,n+2}
   \nonumber\\[2ex]
& & \mbox{}-b_1^2 (n-1) F_{n-2,n+1}-b_1 b_2 (n-2) F_{n-3,n+1}\} .
\end{eqnarray}
 Here
$y_i = k_p(z_i),\ i=1,2$ are the intersections of the path of the pole with
the contour, $\tilde{f}(y)=f(k_p^{-1}(y))$;
 for simplicity, we have omitted the arguments of the functions $F_{n,m}$.

Obviously, the behaviour of $\Delta d_n$ depends
 on the contour chosen.
The general proof of the behaviour of $\Delta d_n$ is, however,
complicated because the function $f(z)$ is unknown. In realistic
situations $f(z)$ is positive and bounded from above by a constant. In the
asymptotic region we  consider the usual asymptotic  expansion
(\ref{Rexp}); which suggests to expand
$f(y)$ in powers of $1/y$.
It is
therefore instructive to consider the case $f(z)=1/y^l$. For $l=0$, we have
\begin{eqnarray}
\label{Fnm}
F_{n,m}|_{\tilde{f}=1}=-\sum_{j=0}^{n} \BC{n}{j} \frac{k_0^{n-j}}{m-j-1}
\times \nonumber\\[2ex]
\times\left(
\frac{1}{(y_2-k_0)^{m-j-1}}-\frac{1}{(y_1-k_0)^{m-j-1}}\right).
\end{eqnarray}
We can now check the different contours; simple examples are (a)
$y_1-k_0=\mbox{const}$ and $y_2-k_0=\mbox{const}$ and (b) the contour
$C_0'$, {\em i.e.\/}, $y_1=\mbox{const}$ and $y_2=\infty $.
It can be seen from eq.\ (\ref{delt}) that in these examples $\Delta
d_n$ diverges when $k_0\to\infty $.
A similar analysis can also be carried out for $f(z)=1/y^l,\ l>0$.
The divergent parts cannot compensate each other, so we are led to the
conclusion that the $\Delta d_n$ are typically divergent. Concluding this
appendix we would like to stress that the contour dependence of $\dnn$ is
quite unacceptable.

\newpage
\section*{Figure Captions}
\vspace{4mm}
\begin{description}
\item[{\bf Fig.1}] The integration contours in the $k$-plane (see Sec.\
3).\\[1mm]
\item[{\bf Fig.2}] The $z$-evolution of the positions of poles inside the
principal strip in the $k$-plane. Two complex conjugate poles ``collide''
 at the origin (for $z=71.024$) and ``scatter'' along the real axis.\\[1mm]
\item[{\bf Fig.3}] The $z$-evolution of the positions of the leading saddle
points. Two real saddle points ``collide'' and set out to the complex plane.
The saddle point near $k_0$ completely disappears for $k_0\to\infty$
 (see eq. (\ref{westphi})). The stars correspond to the positions of
the saddle points at $z=71.024$. \\[1mm]
\item[{\bf Fig.4}] The plot of the real part of the logarithm of
 the integrand in eq. (\ref{newphi}). The values chosen are: $n=4,\ \ln
z=10,\ t=0$ and $k_0=1$.\\[1mm]
\item[{\bf Fig.5}] The contour plot
corresponding to the situation of Fig.4. The resolution is not
sufficient to detect the pole located to the left of the origin.
\end{description} \end{document}